\documentclass[sn-mathphys,Numbered]{sn-jnl}

\usepackage{graphicx}%
\usepackage{multirow}%
\usepackage{amsmath,amssymb,amsfonts}%
\usepackage{amsthm}%
\usepackage{mathrsfs}%
\usepackage{appendix} 
\usepackage{xcolor}%
\usepackage{textcomp}%
\usepackage{manyfoot}%
\usepackage{booktabs}%
\usepackage{algorithm}%
\usepackage{algorithmicx}%
\usepackage{algpseudocode}%
\usepackage{listings}%
\usepackage{hyperref}%
\usepackage{threeparttable}%
\usepackage{tabularx}
\usepackage{bigdelim}
\usepackage[tableposition=top]{caption}
\usepackage{subcaption}
\usepackage{stmaryrd}
\usepackage[framemethod=tikz]{mdframed}
\usepackage{placeins}
\usepackage{booktabs}

\theoremstyle{thmstyleone}%
\theoremstyle{thmstyletwo}%

\theoremstyle{thmstylethree}%

\definecolor{cccolor}{rgb}{.67,.7,.67}

\begin{document}

\title[Article Title]{Detection and classification of vocal productions in large scale audio recordings}


\author*[1,2,4,5]{\fnm{Guillem} \sur{Bonafos}}\email{guillem.bonafos@univ-amu.fr}

\author[1,4]{\fnm{Pierre} \sur{Pudlo}}\email{pierre.pudlo@univ-amu.fr}

\author[1,4]{\fnm{Jean-Marc} \sur{Freyermuth}}\email{jean-marc.freyermuth@univ-amu.fr}

\author[3,4]{\fnm{Thierry} \sur{Legou}}\email{thierry.legou@univ-amu.fr}

\author[2,4]{\fnm{Joel} \sur{Fagot}}\email{joel.fagot@univ-amu.fr}

\author[5]{\fnm{Samuel} \sur{Tronçon}}\email{stroncon@resurgences.eu}

\author[2,4]{\fnm{Arnaud} \sur{Rey}}\email{arnaud.rey@univ-amu.fr}

\affil*[1]{\orgdiv{I2M}, \orgname{Aix Marseille Univ, CNRS}, \orgaddress{\city{Marseille}, \country{France}}}

\affil[2]{\orgdiv{LPC}, \orgname{Aix Marseille Univ, CNRS}, \orgaddress{\city{Marseille}, \country{France}}}

\affil[3]{\orgdiv{LPL}, \orgname{Aix Marseille Univ, CNRS}, \orgaddress{\city{Aix-en-Provence}, \country{France}}}

\affil[4]{\orgdiv{ILCB}, \orgname{Aix Marseille Univ}, \orgaddress{\city{Aix-en-Provence}, \country{France}}}

\affil[5]{\orgdiv{R\'esurgences R\&D}, \orgaddress{\city{Arles}, \country{France}}}


\abstract{We propose an automatic data processing pipeline to extract vocal productions from large-scale natural audio recordings and classify these vocal productions. The pipeline is based on a deep neural network and adresses both issues simultaneously. 
Though a series of computationel steps (windowing, creation of a noise class, data augmentation, re-sampling, transfer learning, Bayesian optimisation), it automatically trains a neural network without requiring a large sample of labeled data and important computing resources. Our end-to-end methodology can handle noisy recordings made under different recording conditions.
We test it on two different natural audio data sets, one from a group of Guinea baboons recorded from a primate research center and one from human babies recorded at home. The pipeline trains a model on 72 and 77 minutes of labeled audio recordings, with an accuracy of 94.58\% and 99.76\%. It is then used to process 443 and 174 hours of natural continuous recordings and it creates two new databases of 38.8 and 35.2 hours, respectively. We discuss the strengths and limitations of this approach that can be applied to any massive audio recording.
}




\maketitle

\section{Introduction}
The manual process of continuous audio recordings to extract and label vocalizations is a complex, tedious and error-prone task. Databases obtained manually are the result of a large and time-consuming task. With an automatic method, we can quickly and cheaply build new massive databases. 
Many continuous-time audio recordings represent significant amounts of data, within which the events of interest are infrequent or even rare. 
Yet these continuous recordings have their own merit: recording an ecosystem without the presence of an experimenter, then automatically extracting vocalizations from it, makes it possible to create new, richer databases. Having a larger number of vocalizations with greater variability would provide domain experts with much more important and relevant information to refine or even challenge repertoire definitions.
In this article, we propose a methodology entirely based on a deep neural network to address the dual challenge of (1) detecting vocalization periods and (2) performing supervised classification of these vocalizations. We need a general workflow, adaptable to find vocalizations of different species, produced in different conditions and different ecosystems. The workflow should be user-accessible, relatively fast and cheap to implement and run. It should require neither massive computational resources nor massive labeled data.

This dual challenge faces numerous issues. Firstly, the vocalization data are necessarily scarce due to manual processing to obtain them, or the limitations of publicly available databases. Secondly, the audio data are captured in uncontrolled environmental conditions, and we have to contend with a variety of background sounds. Thirdly, recording conditions may vary, including different microphone positions or orientations, the use of multiple microphones during recording, and the subject's position relative to the microphone. Fourthly, it is necessary to address the issue of the digital representation of such audio data. Finally, we aim to control the computational cost to maintain reasonable resource usage and facilitate the wider adoption of the proposed procedure.
Traditional detection methods (1) filter the signal, (2) extract a feature vector and (3) use a classification model on this vector \citep{dietrichClassificationBioacousticTime2004, xiaRandomForestClassification2018, nguyenVariationalInferenceBased2018, strisciuglioLearningRepresentationsSound2019}. Neural network avoids these steps learning a hierarchical representation of data automatically \citep{lecunDeepLearning2015}, in an end-to-end manner. It becomes state-of-the-art in bioacoustics and event detection problems \citep{stowellAutomaticAcousticDetection2018}, detecting animal vocal productions when these are in low proportions in massive audio recordings \citep{berglerORCASPOTAutomaticKiller2019}, and despite the noise present in the recordings \citep{oikarinenDeepConvolutionalNetwork2019}. But the limitations remain significant for use on many detection problems \citep{stowellComputationalBioacousticsDeep2022}. These approaches require training databases (and the computational resources that go with them) which are often not available.

The objective of the methodology we propose is to detect variable-duration vocalization segments within a continuous recording and label them according to a pre-labeled training dataset. Our methodology is based on a single deep neural network that handles all the tasks we aim to achieve using raw PCM audio files and a pre-labeled training database. It is described in the second part of this article. We have taken great care in this method to avoid biases. To be able to delineate vocalization segments in the recordings, the machine learning method should be trained on a base that is as representative as possible of the soundscape, including, e.g. biological, geophysical, anthropogenic sounds sounds \citep{pijanowskiWhatSoundscapeEcology2011,berglerORCASPOTAutomaticKiller2019}.
We detail the data required for training and its enrichment, the proposed network architecture, its optimization through data-driven adjustments, the cost function, as well as the transfer learning from a pre-trained network.
Transfer learning has already proved its worth for sound event detection \citep{choiTransferLearningMusic2017, hersheyCNNArchitecturesLargescale2017, palanisamyRethinkingCNNModels2020, aytarSoundNetLearningSound2016}. 
Indeed, we have adopted the representation provided by the YamNet model \citep{TensorFlowHub}. This trained deep neural network is based on a MobileNet architecture \citep{howardMobileNetsEfficientConvolutional2017} and has been trained on the massive AudioSet database \citep{Audioset}. Starting from a mel spectrogram of dimension $96\times 64$, YamNet learns a $1024$-dimensional representation of the audio signal, that is then used to solve a classification problem with $521$ different classes. The massive AudioSet database, composed of more than than 2 billions of 10-seconds audio records, has been used in various bioacoustic tasks \citep{stowellComputationalBioacousticsDeep2022} and for different sound event detection tasks \citep{tenaAutomatedDetectionCOVID192022, patilGearFaultDetection2023}. Additionally, the MobilNet architecture that was implemented to set YamNet, with deep-wise and piece-wise convolution layers, has been designed to be resource-efficient. Relying on YamNet in the middle of our network will ease the resort to our method without placing a heavy burden on computational resources. Moreover, since we keep the weights of YamNet as they are, the fitting on the learning database is simplified.

The third part of this article provides a numerical evaluation of the proposed method using two distinct studies. The first study focuses on recordings of baboons in their habitat at a primate research center. The training database, whether for vocalization detection or classification, is derived from manual labeling of a few recordings captured using the same setup. The second study involves recordings of human infants in their natural environment using recording devices provided to parents. Here, the training database relies on publicly available data concerning infants, which may not possess the same acoustic properties as they originate from other recording conditions. While not an exhaustive evaluation of the proposed method, we draw important conclusions regarding the method's qualities at the end of this part and in the conclusion section. 


\bigskip

\section{Methodology}
This section presents the method we propose for processing these sound recordings.
The recordings are divided into short overlapping frames of $1$ second each. The details of this segmentation and the training databases are provided in Section~\ref{part:data}. Our methodology for processing these frames using a single neural network is described in Section~\ref{part:network}. 
To keep the structure of our network relatively simple, we completely disregard the temporal correlation between successive frames in the continuous recordings as well as in the training databases. While other choices are possible, such as using recurrent networks, training them on medium to large databases can be computationally expensive. 
Furthermore, for sound production detection tasks, the usefulness of using recurrent layers is limited \citep{stowellComputationalBioacousticsDeep2022}.
The adjustment of the network's layers that need to be calibrated on data is described in Section~\ref{part:adjust}. Our objective also requires a step of reconciling predictions on successive frames of the continuous recording, which is also provided in this Section. This reconciliation step allows us to obtain variable-duration vocalization segments and a single labeling per segment.


\subsection{Data}\label{part:data}

To create a suitable training dataset for our dual problem, we need to build two banks of sound recordings. The first bank of audio recordings should consist of labeled vocalization recordings (short or long) without any silence. It aims to address the problem of labeling the detected vocalizations in continuous recordings. These recordings can be manually extracted from the continuous sound recordings and labeled by an expert. Alternatively, they can be sourced from publicly available databases.
The second bank of audio recordings aims to delineate the vocalization segments from the soundscape in our continuous recordings.
The soundscape is the acoustic expression of an ecosystem \citep{pijanowskiWhatSoundscapeEcology2011} or an environment. Hence we highly recommend to build the second bank of background sounds without vocalizations by hand-picking them from the continuous audio recordings. To be as representative as possible of the diversity of noise, they should be picked at various places throughout the whole continuous recordings. Since vocalizations are more seldom than background noises, they should be relatively easy to isolate by hand.
Moreover, once those two banks are compiled, we strongly encourage the resort to relevant data augmentation techniques to enrich these data. We used the ready-to-use library of \citet{mcfeeSoftwareFrameworkMusical2015}, which allows us to multiply by 15 the number of labeled recordings we have in the first bank. From each original file, we shift the pitch by 5 values linearly spaced within $(-4,4)$, we stretch the speed by 5 values logarithmically spaced within $(0.81,1.23)$, and we add 5 background noise from the second bank.

To standardize the audio recordings, we assume that they are all available as a single-channel audio signal, sampled at 16 kHz in PCM format, for example as WAV files.
The pulse modulation signal is translated to be on a scale between $-1$ and $+1$. Through windowing, the signal is divided into frames of one second each, with an overlap of $80\%$. This divides a recording of $T$ seconds into $5T$ frames. The position of each frame along the continuous recordings should be saved. The same process is also applied to the recordings of the two banks, yet saving their position is useless. 
Note that a time window of one second is consistent with the problem with which we are dealing. It allows us to quickly discard noise segments from the data to be analyzed. And one second seems to be a good compromise, sufficient to encompass most vocalizations but not too large to be easily processed.

Whether it is the frames obtained from the decomposition of the continuous recording to be analyzed or from the two banks, we use the following conventions in the notations: $x$ represents a one-second audio signal, $y \in \{0, 1\}$ is a vocalization indicator, and $z \in \{1, 2, \ldots, K\}$ is the label of the vocalization. Note that if $x$ is not a vocalization, $y = 0$ and $z$ takes an arbitrary value.
At this stage, we can aggregate the frames coming from the two banks as a single table of triplets $(x_i, y_i, z_i)$, where $i = 1, 2,\ldots$.
As usual in machine learning, the table should be divided in three parts: a training dataset ($\approx 60\%$), a validation and a test dataset ($\approx 20\%$ each).

To serve as input to the YamNet deep neural network \citep{TensorFlowHub}, these one-second frames need to be represented in the time-frequency domain using a log-mel spectrogram. The log-mel spectrogram mimics the sensitivity of the human ear to frequency and amplitude differences. This transformation is a frequent pre-processing step of audio signal to input them into deep learning models. The following steps are required to obtain an input of dimension $96\times 64$: (1) apply the Short-Time Fourier Transform (STFT) with a window size of 25 ms, a hop size of 10 ms, and a Hann window function, (2) segment the spectrum into 64 mel bins spanning the frequency range of $125$ \-- $7500$ Hz, and (3) apply a logarithmic scaling. 
This process can be accomplished using one layer of a neural network. We have made this choice in our code for the sake of efficiency in our pipeline.

\subsection{Network architecture}\label{part:network}

\begin{table}[t]
    \centering
    \fontsize{8pt}{8pt}\selectfont
        \caption{Network Architecture}
        \label{tab:model_architecture}
            \begin{tabular}{c*{2}c}
            \toprule 
            & Type & Input Size \\
            \midrule
            & Log-scaled Mel Spectrogram & $16000\times 1$ \\
            \cmidrule{2-3}
            \multirow{5}{*}{\rotatebox[origin=c]{90}{YamNet}}& Convolution & $96\times 64\times 1$ \\
            \addlinespace
            & $\vdots$ & $\vdots$  \\
            \addlinespace
            & Average Pooling & $3\times 2\times 1024$ \\
            \cmidrule{2-3}
            \multirow{12}{*}{\rotatebox[origin=c]{90}{Detection}} & Fully Connected & $1\times 1\times 1024$ \\
            & Batch-Normalization & \\
            & Drop-out & \\
            \addlinespace
            & Fully Connected & $1\times 1\times m_{1}^d$ \\
            & Batch-Normalization & \\
            & Drop-out & \\
            \addlinespace
            & $\vdots$ & $\vdots$  \\
            \addlinespace
            & Fully Connected & $1\times 1\times m_{(\ell^d-1)}^d$ \\
            & Batch-Normalization & \\
            & Drop-out & \\
            \addlinespace
            & Sigmoid & $1\times 1\times m_{\ell^d}^d$ \\
            \cmidrule{2-3}
            \multirow{12}{*}{\rotatebox[origin=c]{90}{Classification}} & Fully Connected & $1\times 1\times 1024$\\
            & Batch-Normalization & \\
            & Drop-out & \\
            \addlinespace
            & Fully Connected & $1\times 1\times m_{1}^c$\\
            & Batch-Normalization & \\
            & Drop-out & \\
            \addlinespace
            & $\vdots$ & $\vdots$  \\
            \addlinespace
            & Fully Connected & $1\times 1\times m_{(\ell^c-1)}^c$\\
            & Batch-Normalization & \\
            & Drop-out & \\
            \addlinespace
            & Softmax & $1\times 1\times m_{\ell^c}^c$\\
            \addlinespace
            \bottomrule
            \end{tabular}
    \smallskip
    \scriptsize
        \footnotesize $\ell^d, \ell^c \in \llbracket 1, 6 \rrbracket$, learned on the data. \\
        $m_{i}^d \in \llbracket 32, 1024 \rrbracket$, learned on the data, for $i=1,\ldots,\ell^d$ \\
        $m_{i}^c \in \llbracket 32, 1024 \rrbracket$, learned on the data, for $i=1,\ldots,\ell^c$ 
\end{table}

We rely on the YamNet network to move from the mel spectrogram of dimension $96 \times 64$ to a representation of the audio frame in dimension $1024$ that is more relevant for the dual problem at hand. The first layer that computes the mel spectrogram, as well as the layers transferred from YamNet are not fitted on the data described in Section~\ref{part:data}. 

The dual problem we are trying to solve begins with the prediction of $(y, z)$ given the observation $x$ of a one-second frame.  From a mathematical point of view, the problem can be reduced to the prediction of $K+1$-classes' problem by predicting $t=yz\in\{0,1,2,\ldots, K\}$: we simple add a non-vocalization class (labeled by $0$) to the classes of vocalization. Yet we expect that the difference between a silence or a noise from the soundscape and a vocalization is much larger than the subtle difference between two classes of vocalizations. 
Resolving the detection problem is thus a simpler learning task compared to the classification of vocalizations. We might expect a high error rate in the subsequent classification of detected vocalizations. Yet it is crucial to maintain strict control over the error rate in vocalization detection since our primary focus is on finding the seldom vocalization in continuous recordings. Furthermore, using the 1024-dimensional representation of YamNet, the detection problem should not be approached in the same way as the classification problem: the relevant coordinates and/or the manner in which to use these coordinates to solve both problems are likely to be different. Based on these considerations and supported by preliminary numerical results (not presented in this article), we have chosen to treat the detection and classification problems separately while developing a single neural network.

Instead of directly predicting $(y, z)$, we construct a network that outputs estimates $\hat p(x)$ and $\hat q(x)$ of the posterior probability vectors:
\begin{align*}
p_k(x) &= \mathbb P(y=k|x), \quad k=0,1
\\
q_k(x) &= \mathbb P(z=k|x), \quad k=1,\ldots,K. 
\end{align*}
The value of $z$ indicates the class of the vocalization and is only meaningful if $x$ is such a recording. Hence, our loss function is
\begin{equation}\label{eq:loss}
L\Big( (y,z), (\hat p, \hat q)\Big) =
 - (1-y)\log \hat p_0 - y \log \hat p_1 -
y \bigg( \sum_{k=1}^K \mathbf1\{z=k\}\log \hat q_k\bigg)
\end{equation}
where the second cross-entropy term plays a role only if $y=1$, i.e., only if it is a vocalisation.

For all those reasons, we add two separated modules on top of the last layer of YamNet. They aim at estimating $p(x)$ and $q(x)$ respectively. Both need to be trained on data and are composed of fully connected layers with Parametric Rectified Linear Unit activation function \citep{heDelvingDeepRectifiers2015}. The weights of the two modules are initialized following the initialization scheme proposed by \citet{heDelvingDeepRectifiers2015}, which should facilitate the convergence of the model \citep{kumarWeightInitializationDeep2017}. 
The number of layers of each module is between 1 and 6, the number of nodes by layer between 32 and 1024.
We rely on a regularization strategy to avoid problem of over-fitting: a batch normalization \citep{ioffeBatchNormalizationAccelerating2015} is computed after each layer, as well as drop-out and a max constraint on the norm of the weights \citep{srivastavaDropoutSimpleWay2014}.
The activation function of the last layer of each module is either a sigmoid or a softmax function to get the desired posterior probabilities. 
The resulting architecture is described in Table~\ref{tab:model_architecture}.

\subsection{Fit on the data}\label{part:adjust}
Gradient based algorithm show the advantage of our architecture. To learn from our training dataset, we minimize the loss in Equation~\eqref{eq:loss} with the NAdam algorithm \citep{dozatIncorporatingNesterovMomentum2016, kingmaAdamMethodStochastic2017}. 
Many layers of our network are frozen and transferred. Yet the layers of the two modules should be fitted to the data. 
When the input data is a non-vocalization frame, the loss reduces to the cross-entropy of the detection module's output $\hat p(x)$. Its gradient with respect to the network weights to be adjusted is therefore zero on the weights of the classification module. On the other hand, when the input data is a classified vocalization frame, the loss is the sum of two cross-entropies, each computed on the output of one of the modules, either $\hat p(x)$ or $\hat q(x)$. In this case, the gradient decomposes into the sum of two vectors: the first vector only affects the weights of the detection module, and is obtained by taking the gradient of the detection cross-entropy;  the second vector only affects the weights of the classification module, and is derived from the gradient of the classification cross-entropy.
Thus, the adjustment of both modules can be done simultaneously, without the improvement of one module degrading the performance of the other module. However, if we had relied on a single module to solve the single classification problem with $(K+1)$ classes by adding a non-vocalization class to the $K$ vocalization classes, this observation would no longer hold. As the detection problem is simpler than the vocalization classification problem (see Section~\ref{part:network}), the single network struggles to adjust to the dual problem. The two-module solution is a way to avoid bias due to the fact that detection is simpler than classification of vocalizations.

To avoid other bias, we need to take care of the way training frames enter in the NAdam algorithm, and deal with possible unbalanced training data. The solution we propose is as follows. Each batch is composed of $N_\text{batch}$ triplets $(x_i, y_i, z_i)$ drawn from the training dataset with replacement. Each triplet is drawn as follow. First, we draw $y^\dag$ uniformly over $\{0,1\}$. If $y^\dag=0$, we draw the triplet at random among the non-vocalization triplets, namely $\{(x_i, y_i, z_i):y_i=y^\dag\}$. Otherwise, $y^\dag=1$, we draw $z^\dag$ uniformly over $\{1,\ldots, K\}$ and pick a triplet at random among the vocalization triplets of class $z^\dag$, namely $\{ (x_i,y_i,z_i):y_i=y^\dag, z_i=z^\dag\}$. Thus, on average, half of each batch is composed of non-vocalization frames; and among the vocalization frames, the $K$ classes are balanced. Since the data seen by NAdam is an infinite flow of triplets drawn at random from the training dataset, we define an epoch as a set of $N_\text{epoch}=(K+1)N_\text{max}/N_\text{batch}$ batches, where $N_\text{max}$ is the size of the set of the largest class of vocalization and $N_\text{batch}$ the size of the batch.
Along the NAdam algorithm, the learning rate is decreased by a factor of $0.2$ if the validation loss has not decreased after $5$ epochs. And the whole algorithm is stopped after $20$ epochs without validation error decrease.

The hyperparameters of the two modules are calibrated on the validation dataset using a Bayesian optimization scheme \citep{brochuTutorialBayesianOptimization2010, snoekPracticalBayesianOptimization2012, shahriariTakingHumanOut2016} which minimizes the validation computed with the loss given in Equation~\eqref{eq:loss}. The parameters of the NAdam algorithm, as well as the parameters that defines the precise architecture of the two modules are calibrated with this method, are given in Table~\ref{tab:param_Bayesian}.

\begin{table}[tbp]
\caption{Hyper-parameters of the model}
\label{tab:param_Bayesian}
\centering {\fontsize{9pt}{9pt}\selectfont
\begin{tabular}{cc} 
\toprule
Hyper-parameters & Research Space \\
\midrule
$p_\text{drop-out}$ (dropout within both modules)& $\left[0.1,0.9\right]$ \\
$c_\text{norm}$ (batch-normalization within both modules) & $\llbracket 0,8 \rrbracket$\\
$\alpha$ (learning-rate) & $\left[1e-10,1e-2\right]$ \\
$\beta_1$ (decay-rate of the moving average of the gradient) & $\left[0,0.9\right]$ \\
$\beta_2$ (decay-rate of the moving average of the squared gradient) & $\left[0.99,0.9999\right]$ \\
$\ell^d$ (number of hidden fully connected layers) & $\llbracket 1, 6 \rrbracket$ \\
$\ell^c$ (number of hidden fully connected layers of the classification module) & $\llbracket 1, 6 \rrbracket$ \\
$m_i$, for $i=1,\ldots,\ell^d\text{ and }\ell^c$ (number of nodes per layer) & $\llbracket 32, 1024 \rrbracket$ \\
\bottomrule
\end{tabular}
}
\end{table}

\subsection{Vocalization delineation and classification} \label{part:prediction}

The use of the trained network to delineate a vocalization period on a continuous audio recording as well as its classification needs to be explained. We need a conciliation procedure that reintroduce the temporal dependency of our overlapping frames that was lost by the network. More precisely, we need a rule that allows errors in the detection of vocalization frames, and in the classification of detected vocalization frames. To this aim, we design the procedure given below with the following rules. First, a vocalization period can include period of time of length less than one second where vocalizations have not been detected by the network. Second, to aggregate the class predictions, we use a majority vote.

As before, the recording to be analyzed of length $T^\star$ seconds should be divided into frames of $1$s with an overlap of $80\%$. Let us denote $x_t^\star$, $t=1,\ldots, 5T^\star$ the $t$-th frame of this recording.
Using the trained network, we can compute for each $t$ the maximum a posteriori:
\begin{align*}
    \hat{y}^\star_t &= \begin{cases}
        1 & \text{if } \hat p_1(x_t^\star)>0.5,
        \\
        0 & \text{otherwise}
    \end{cases}&
    \quad \quad
    \hat{z}^\star_t &= \text{argmax}_k \hat q_k(x^\star_t). 
\end{align*}

In order to delineate a vocalization segment based on the predicted values $\hat y^\star_t$, $t=1,\ldots$, we introduce a equivalence relation on the set $\mathcal T_1 = \{t: \hat{y}^\star_t=1\}$ as follows. We say that $t\sim t'$ if and only if there exists an increasing sequence $t_0,\ldots, t_N\in \mathcal T_1$ such that $t_0=t$, $t_N=t'$ and $t_{i+1}-t_{i}\le 5$. It means that the audio segment starting with the frame at position $t$ and ending with the frame at position $t'$ is composed of frames that are predicted as vocalizations, expect on small time periods that last less than one second. The equivalence classes of this relation are easy to determine along time. The position of the starts and ends of the vocalization segments along the continuous recording are then given by the starts and the ends of the equivalence classes of this equivalence relation.

Once a vocalization segment is delineated, we have to predict its class. To this aim, we use the predicted classes $\hat{z}^\star_t$ of the frames that compose the segment. Considering our loss function given in Equation~\eqref{eq:loss}, the predicted classes are reliable only on the frames that have been detected as vocalizations. Among these reliable predictions, a majority vote allows us to determine the class of the vocalization segment.

\section{Experimental Validation} \label{part:experimental}

The proposed pipeline has been tested in two different studies: a first study dealing with baboon vocalizations and a second one dealing with human baby vocalizations. The first study is a bioacoustic problem that aims at collecting vocalizations of Guinea baboons (\textit{Papio papio}). The second study is a developmental psycho-acoustic problem and is focused on the vocalizations of human infants between 0 and 12 months of age. For each study, the output is a large-scale database of labeled vocalizations. 

The data of the two studies are described in Section~\ref{part:continuous_data}.

\subsection{From audio recordings to data banks for our method}\label{part:continuous_data}

Both studies aim at capturing vocalizations that are not provoked by the experimental setup, using continuous and possibly daylong recordings. However, the recording conditions are quite different. These two studies provide a first example of the diversity of situations to which our method can adapt, as each of them contains different sound events other than vocalizations, as well as different background noises that interfere with the vocalizations. 

In the baboon study, we recorded continuously during approximately one month a group of 25 Guinea baboons (\textit{Papio papio}) from the CNRS primatology center of Rousset-sur-Arc (France). The group lives in semi-liberty in a large rectangular enclosure outdoors. Ethical agreement (\# 02054.02) was obtained from the CEEA-14 for experimental animal research to conduct audio recordings of the baboons' vocalizations. Two microphones are placed at two corners of the enclosure. In addition to the baboons' vocalizations, the sound environment is composed of climatic events (wind, rain), the presence of other nearby animal species (sheep, birds), and human activities (people around the enclosure, cars on the nearby highway, planes, etc.). One month of recording leads to a tremendous amount of data: after removing night recordings when baboons are at sleep inside a room (from 9 pm to 7 am), there is a total of 460 files representing 443 hours of recording (i.e., 1 595 018.24 seconds). 

In the human study, we collected recordings from two human babies at home from birth to their first birthday, at a rate of three days per month. An ethical agreement (\# 2019-12-12-005) was obtained from the ethics committee of Aix-Marseille University as well as a declaration of conformity from the CNIL (\# 2222631 v 0) for experimental research on humans in order to make audio recordings of human baby vocalizations. All parents gave their informed consent for inclusion before their inclusion in the study. The records were done by the parents at different moments of the days and nights. The parents were instructed to start and stop themselves the recordings. Although less noisy than the baboon environment, the recordings are composed of a lot of heterogeneous sources of sounds: TV, radio, domestic works, parents, other children. In total, the records represent 174.15 hours (626 940 seconds) for the two children.

Both studies had some uncontrolled recording conditions. In the baboon study, the microphone had a fixed position and the signal source is mobile. The monkeys move and vocalize from various location into different direction. In the human study, the microphone is constantly changing position. With each new recording, the parents place the microphone in a position that may be different from the source, the baby. 

In addition to the continuous sound recordings to be analyzed, our method is based on two recording banks for each study: a first bank of non vocalization recordings, and a second bank of labeled vocalization recordings. In both studies, the first bank that helps to delineate vocalizations was extracted from the continuous recordings, as proposed in Section~\ref{part:data}. In the baboon study, we listened to a total amount of 7 hours of these recordings from which we removed the vocalizations. In the human study, we used the same method on a total amount of 5 hours of recordings. In both cases, we took care to get a bank as representative as possible of the various sound events and noises: the excerpts we listened to were chosen at different dates and times of the day and came from different families. For the baboon study, it represents 355.62 minutes. For the baby study, it represents 104.56 minutes.

The bank of labeled vocalization recordings were constructed differently in both studies. In the baboon study, we had from a previous study \citep{boeEvidenceVocalicProtoSystem2017} a total amount of 72.49 minutes of  labeled vocalization, divided into 6 classes: bark, copulation grunt, grunt, scream, wahoo, and yak. These baboon vocalizations came from the same group of baboons and from the same experimental setup. In contrast, the baby vocalizations came from a public database \citep{cychoszBabbleCorCrosslinguisticCorpus2019}, based on daylong audio recordings of 49 children (1–36 months) from five different languages and cultural backgrounds that were annotated by citizen scientists. This public database gave us a bank of labeled vocalizations that represents a total amount of 77.03 minutes of recordings, divided into 5 classes: canonical, crying, junk, laughing, non-canonical. In both studies, the classes in the vocalization bank were unbalanced. More details on the composition of the two banks in both studies are given in the Supplementary Materials (Table \ref{tab:effectif_papio} and \ref{tab:effectif_baby})

\subsection{Performance of our deep learning architecture}

Before presenting the final results of the proposed methodology, we start by analyzing the outputs of our network on the 1-second frames from the test datasets of both studies: Table~\ref{tab:metrics} provides different evaluations of the network's performance on our dual problem, calculated using standard metrics. Detailed confusion matrices for the detection problem as well as the classification problem in both studies are given in Supplementary Materials (Figures \ref{fig:mc_papio} and \ref{fig:mc_bebe}). At first glance, the results are relatively similar, suggesting that our pipeline can be used successfully in different types of studies. 
In particular, the performances in the detection problem measured with the precision and recall metrics indicate that we succeeded in our primary goal of detecting correctly the vocalization frames. 

As can be seen in Table~\ref{tab:metrics}, the detection problem is resolved more satisfactorily in the study on human infants than in the study on baboons. We can propose two explanations for this discrepancy. Firstly, the quality of the recordings of baby vocalizations is much better than that of baboons: they were made in a much quieter place, at quieter times of the day, and with a microphone likely closer to the source, whereas baboons move in a noisier and larger environment. Particularly, on windy days (Mistral), the recordings of the baboon study are of very poor quality. Secondly, the 1024-dimensional representation provided by YamNet is likely more suited to detect human vocalizations than baboon vocalizations. Indeed, this representation was learned on the Audioset database to solve a classification problem with 521 classes. This massive database contains numerous antropophonic sounds distributed across several classes. But vocalizations of animals are much rarer and distributed across coarser classes.

On the other hand, we can see in Table~\ref{tab:metrics} that the classification problem is resolved more satisfactorily in the baboon study even if it is a 6-classes problem whereas the baby vocalizations are divided in 5 classes only. Even though YamNet was not originally designed to understand the differences between baboon vocalizations, the results of our methodology are relatively good at classifying them. Moreover, the distinction between the classes of infant vocalizations is likely more subtle than the differences between the classes of baboon vocalizations. The study on infants marks the beginning of an investigation into language development. In the early months, the infants are still in the process of learning this classification for themselves, and the differences between the vocalizations produced can sometimes be subtle. A more closer look at the confusion matrices given in Supplementary Materials shows that the major part of errors in the baboon study comes from rather similar classes: ‘‘copulation grunt'' which is a specific class in our study can be seen as a specific type of ‘‘grunt'' and a ‘‘bark'' vocalization share common features with a ‘‘yahoo''.

We processed the continuous audio recordings for both problems on a laptop  with a single GPU, on which Tensorflow \citep{tensorflow2015-whitepaper} was able to train and run the deep network.
The 460 files representing 443 hours of continuous recording of the baboon have been loaded, segmented, and classified in 9 hours 28 minutes. The 261 files representing 174 hours of continuous recording for the two human babies have been loaded, segmented, and classified in 9 hours 44 minutes.

\begin{table}
\caption{Performance of the deep learning in the baboon and human studies}
\label{tab:metrics}
\centering
\fontsize{8pt}{8pt}\selectfont
\begin{tabular}{cccc}
\toprule & & \multicolumn{2}{c}{Data set} \\
\cmidrule{3-4}
& & Baboon & Baby \\
\midrule
& Loss (from equation \ref{eq:loss}) & 0.12 & 0.07 \\
\midrule
\multirow{5}{*}{Detection} & Cross-entropy & 0.04 & 0.01 \\
\addlinespace
& Accuracy & 94.58 & 99.76 \\
\addlinespace
& AUC & 0.94 & 0.99 \\
\addlinespace
& Precision & 82.68  & 99.33\\
\addlinespace
& Recall & 90.28 & 99.74\\
\midrule
\multirow{2}{*}{Classification} & Cross-entropy & 0.23 & 0.26 \\
\addlinespace
& Accuracy & 48.92 & 39.96 \\
\bottomrule
\end{tabular}

\end{table}

\subsection{New large-scale databases of vocalizations}

Once the model has been trained on the labeled data, we can use it on the massive continuous data to extract the moments of vocalization and create two new large-scale data sets. We can measure the amount of data extracted and the time to do it.

Two new databases are constructed from the continuous recordings processed by the model through our pipeline. The new human baby database represents 35.20 hours of records. The new baboon database represents 38.75 hours of records. Table \ref{tab:baby_total_voc_year} and \ref{tab:papio_total_voc_month} respectively, summarizes the distribution for each class, for each data.

We have made the data extracted by our model from continuous baboons recordings freely accessible on \href{https://zenodo.org/record/7963124}{https://zenodo.org/record/7963124}. This includes the labeled database used for training, as well as the vocalizations detected during the month and a csv file summarizing, for each vocalization, its duration, probability and vote for each class. A typical day was also made available, to give everyone the chance to test the model on this typical example (it was not possible to make more than one day of continuous recording available for legal reasons). The code is accessible on \href{https://gitlab.com/papers4375727/detection-and-classification-of-vocal-productions.git}{https://gitlab.com/papers4375727/detection-and-classification-of-vocal-productions.git}, with an exemple with this day to reproduce the work.
Results for baby recordings are not accessible for legal reasons.

\bigskip

\section{Conclusion and Discussion}

The goals of the pipeline were to quickly process and classify hundreds of hours of audio, with as few errors as possible, minimizing information loss, through an end-to-end pipeline with no engineering steps, so that it can be reused in different situations. In addition, the pipeline had to adapt to various environmental sound classification problems, with little labeled data for learning. 

Our two-module architecture, together with the care taken with the training set, enables us to achieve high scores on precision and recall in the vocalization detection problem. This was the primary objective of our methodology, and we can consider that it has been achieved without mobilizing massive computing resources, thanks to the transfer of YamNet. Even on vocalizations with which YamNet is unfamiliar, such as those of baboons, the detection scores (precision and recall) remain very high, showing that our method is capable of attacking a certain diversity of species. 

The two studies of Section~\ref{part:experimental} show that YamNet has sufficient generality to tackle a wide class of problems similar to the one addressed in this work. AudioSet is massive enough to learn a representation which distinguish between one type of signal and another, between a species' vocalization and the rest of its soundscape.
The two studies show that our method is robust to a variety of complicated recording conditions, and generic enough for use in a variety of contexts, species,\ldots 
The limits are certainly in the frequency range to which YamNet is sensitive. This range is similar to that of the human ear; it does not allow us to deal with species such as bats, for example, which emit sounds in the high treble or ultra-sounds. The representation transferred by YamNet is undoubtedly not the most appropriate for tackling a problem without adopting an anthropocentric stance, since it is based on a representation designed for the human ear \citep{pratAnimalsHaveNo2019}.

Thanks to our two-module architecture, we can handle a second problem simultaneously with detection, such as the classification of vocalizations, without degrading detection scores. Classification scores are less satisfactory, but this problem, like many learning problems on vocalizations, requires the learning model to be able to distinguish more subtle differences. Probably, our training bases were too limited, representing a total of one hour of recording in which our classes were unbalanced. 

The conciliation procedure we have introduce in Section~\ref{part:prediction} to reintroduce the temporal dependency is quite rough. We discarded other attempts that introduced too many computational burden for the output provided. Yet, we have lost the uncertainty measure provided by the network through $\hat p(x)$ and $\hat q(x)$. It would be interesting to develop a probabilistic method capable of performing this reconciliation, without weighing down our pipeline numerically.

\section*{Statements and Declarations}

\bmhead{Funding}
This work was carried out in a collaboration between the CNRS, Aix-Marseille University and Résurgences R\&D around the CIFRE PhD n$^\text{o}$215582 with the support of the ANRT, within the Labex BLRI (ANR-11-LABX-0036) and the Institut Convergence ILCB (ANR-16-CONV-0002). It also benefited support from the French government, managed by the French National Agency for Research (ANR) and the Excellence Initiative of Aix-Marseille University (A*MIDEX). This work was also supported by the CHUNKED ANR project (ANR-17-CE28-0013-02). The funders had no role in study design, data collection and analysis, decision to publish, or preparation of the manuscript. 

\bmhead{Acknowledgments}
We are grateful to Rosic Ferry-Huiban and Myriam Sabatier for their help in labeling the baboon database. 

\bmhead{Conflict of interest}
The authors have no relevant financial or non-financial interests to disclose.

\bmhead{Ethics approval}
Ethical agreement (\# 02054.02) was obtained from the CEEA-14 for experimental animal research to conduct audio recordings of the baboons' vocalizations.   
An ethical agreement (\# 2019-12-12-005) was obtained from the ethics committee of Aix-Marseille University as well as a declaration of conformity from the CNIL (\# 2222631 v 0) for experimental research on humans in order to make audio recordings of human baby vocalizations.

\bmhead{Availability of data and materials}
Data set created on the continuous baboons recordings is accessible on \href{https://zenodo.org/record/7963124}{https://zenodo.org/record/7963124}, with the training set used for this study and two hours as exemple.    
Data set created on the continuous baby recordings is not accessible for legal reasons. The labeled data for the babies came from a public database named Babblecor \citep{cychoszBabbleCorCrosslinguisticCorpus2019} which can be found on \href{https://osf.io/rz4tx/}{https://osf.io/rz4tx/}

\bmhead{Code availability}
The pipeline proposed in this paper is available on GitLab at  \href{https://gitlab.com/papers4375727/detection-and-classification-of-vocal-productions.git}{https://gitlab.com/papers4375727/detection-and-classification-of-vocal-productions.git}. The repository contains the implementation of the algorithms described in the methods section, along with detailed documentation on how to use the code.

\bigskip

For the purpose of Open Access, a CC-BY\footnote{https://creativecommons.org/licenses/by/4.0/} public copyright licence has been applied by the authors to the present document and will be applied to all subsequent versions up to the Author Accepted Manuscript arising from this submission.

\medskip

\begin{mdframed}[outerlinecolor=black,outerlinewidth=2pt,linecolor=cccolor,middlelinewidth=3pt,roundcorner=10pt, nobreak=true]
 Distributed under a Creative Commons Attribution | 4.0 International licence.
 \smallskip
  \begin{center}
    \includegraphics[scale=1]{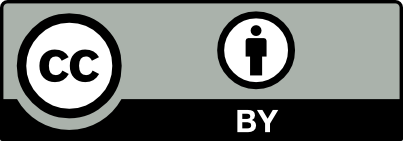}
  \end{center}
\end{mdframed}


\bibliography{papiovoc}

\appendix
\makeatletter
\renewcommand \thesection{S\@arabic\c@section}
\setcounter{section}{0}

\section{Supplementary Materials}\label{part:supp_mat}


\begin{figure}[htb]
\centering
\begin{subfigure}[b]{0.4\textwidth}
\centering
\includegraphics[width=\textwidth]{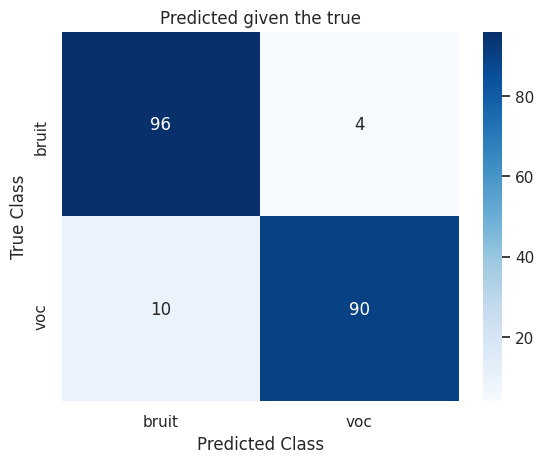}
\caption{Detection of vocalizations against noise.}
\label{fig:mc_bin_papio}
\end{subfigure}
\hfill
\begin{subfigure}[b]{0.4\textwidth}
\centering
\includegraphics[width=\textwidth]{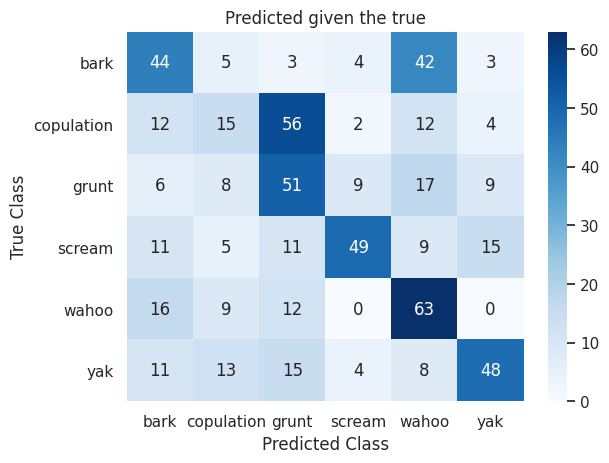}
\caption{Classification of the detected vocalizations.}
\label{fig:mc_multi_papio}
\end{subfigure}
\caption{Confusion matrices for the \textit{baboon} data.}
\label{fig:mc_papio}
\end{figure}

\begin{figure}[htb]
\centering
\begin{subfigure}[b]{0.4\textwidth}
\centering
\includegraphics[width=\textwidth]{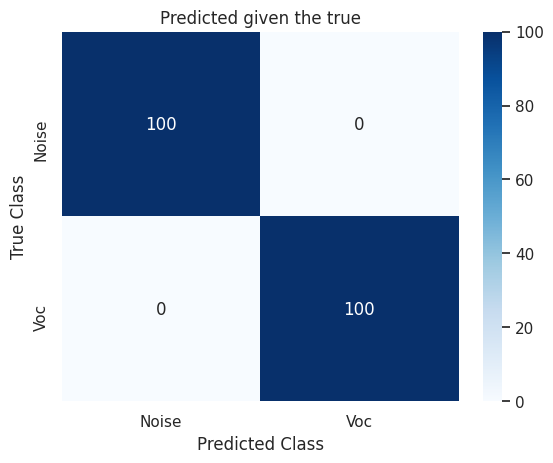}
\caption{Detection of vocalizations against noise.}
\label{fig:mc_bin_bebe}
\end{subfigure}
\hfill
\begin{subfigure}[b]{0.4\textwidth}
\centering
\includegraphics[width=\textwidth]{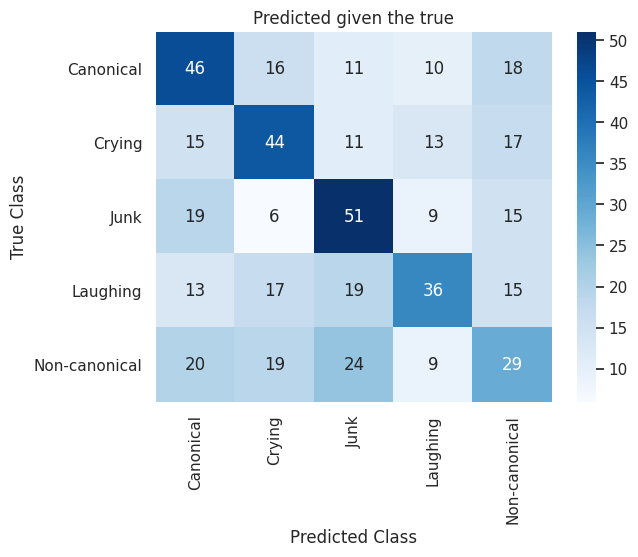}
\caption{Classification of the detected vocalizations.}
\label{fig:mc_multi_bebe}
\end{subfigure}
\caption{Confusion matrices for the \textit{human baby} data.}
\label{fig:mc_bebe}
\end{figure}



\begin{table*}[htb]
\caption{Total and per partition distribution of the baboon labeled data set.}
\label{tab:effectif_papio}

\begin{subtable}{\textwidth}
\centering
\subcaption{Initial sample size, number of records and duration for each vocalization class, for the baboon data set.}
\label{tab:effectif_initial_papio}
\begin{tabularx}{\textwidth}{X *{6}c}
\toprule
& \multicolumn{6}{c}{Classes}\\
\cmidrule{2-7}
& Bark & Copulation & Grunt & Scream & Wahoo & Yak \\
\midrule
Number of records & 269 & 68 & 502 & 414 & 97 & 60\\
\addlinespace
Duration in seconds & 192.1 & 132.1 & 2007.5 & 1695.8 & 146.4 & 175.5 \\
\addlinespace
Mean record time (secs) & 1.3 & 3.0 & 6.7 & 7.0 & 2.4 & 4.9 \\
\bottomrule
\end{tabularx}
\end{subtable}

\hspace{\fill}

\begin{subtable}{\textwidth}
\centering
\vspace{20pt}
\subcaption{Number of records and their duration per vocalization class for the training set of the baboon data set}
\label{tab:effectif_train_papio}
\begin{tabularx}{\textwidth}{X *{7}c}
\toprule
& \multicolumn{7}{c}{Classes}\\
\cmidrule{2-8}
& Bark & Copulation & Grunt & Scream & Wahoo & Yak & Noise \\
\midrule
Numbers of records & 152 & 44 & 302 & 244 & 51 & 36 & 82\\
\addlinespace
Duration in seconds & 98.5 & 84.3 & 1252.9 & 951.6 & 105.3 & 99.1 & 13884.7\\
\addlinespace
Mean record time (secs) & 0.7 & 1.9 & 4.2 & 3.9 & 1.7 & 2.8 & 169.3\\
\bottomrule
\end{tabularx}
\end{subtable}

\end{table*}


\begin{table*}
\ContinuedFloat
\caption{Total and per partition distribution of the baboon labeled data set.}

\begin{subtable}{\textwidth}
\centering
\vspace{20pt}
\subcaption{Number of records and their duration per vocalization class for the validation set of the baboon data set}
\label{tab:effectif_val_papio}
\begin{tabularx}{\textwidth}{X *{7}c}
\toprule
& \multicolumn{7}{c}{Classes}\\
\cmidrule{2-8}
& Bark & Copulation & Grunt & Scream & Wahoo & Yak & Noise\\
\midrule
Numbers of records & 54 & 11 & 101 & 89 & 20 & 12 & 20\\
\addlinespace
Duration in seconds & 36.1 & 20.7 & 362.0 & 340.2 & 20.3 & 54.2 & 3379.0\\
\addlinespace
Mean record time (secs) & 0.2 & 0.5 & 1.2 & 1.4 & 0.3 & 1.5 & 169.0 \\
\bottomrule
\end{tabularx}
\end{subtable}

\hspace{\fill}

\begin{subtable}{\textwidth}
\centering
\vspace{20pt}
\subcaption{Number of records and their duration per vocalization class for the testing set of the baboon data set}
\label{tab:effectif_test_papio}
\begin{tabularx}{\textwidth}{X *{7}c}
\toprule
& \multicolumn{7}{c}{Classes}\\
\cmidrule{2-8}
& Bark & Copulation & Grunt & Scream & Wahoo & Yak & Noise\\
\midrule
Numbers of records & 63 & 13 & 99 & 81 & 16 & 12 & 24\\
\addlinespace
Duration in seconds & 57.5 & 27.2 & 392.6 & 404.0 & 20.7 & 22.2 & 4073.4\\
\addlinespace
Mean record time (secs) & 0.4 & 0.6 & 1.3 & 1.7 & 0.3 & 0.6 & 169.7 \\
\bottomrule
\end{tabularx}
\end{subtable}

\end{table*}

\begin{table}[htb]
\caption{Seconds of vocalization for baboons, for each class, over the month.}
\begin{tabularx}{\textwidth}{*{6}X}
\toprule \multicolumn{6}{c}{Classes} \\
\cmidrule{1-6}
Bark & Copulation & Grunt & Scream & Wahoo & Yak \\
\midrule
3 197 & 8 455 & 50 679 & 35 070 & 23 026 & 19 080 \\
\bottomrule
\end{tabularx}
\label{tab:papio_total_voc_month}
\end{table}

\begin{table*}[htb]
\centering
\caption{Total and per partition distribution of the human baby labeled data set}
\label{tab:effectif_baby}

\begin{subtable}{\textwidth}
\centering
\caption{Initial sample size, number of records and their duration per vocalization class for the \textit{human baby} data set.}
\label{tab:effectif_initial_bebe}
\begin{tabularx}{\textwidth}{X *{5}c}
\toprule
& \multicolumn{5}{c}{Classes}\\
\cmidrule{2-6}
& Canonical & Crying & Junk & Laughing & Non-Canonical \\
\midrule
Numbers of records & 1826 & 823 & 4974 & 241 & 5606 \\
\addlinespace
Duration in seconds & 677.9 & 297.6 & 1665.1 & 87.1 & 1894.0 \\
\addlinespace
Mean record time (secs) & 0.4 & 0.4 & 0.4 & 0.4 & 0.4 \\
\bottomrule
\end{tabularx}
\end{subtable}

\hspace{\fill}

\begin{subtable}{\textwidth}
\centering
\vspace{20pt}
\subcaption{Number of records and their duration per vocalization class for the training subset of the \textit{human baby} data set}
\label{tab:effectif_train_bebe}
\begin{tabularx}{\textwidth}{X *{6}c}
\toprule
& \multicolumn{6}{c}{Classes}\\
\cmidrule{2-7}
& Canonical & Crying & Junk & Laughing & Non-Canonical & Noise \\
\midrule
Numbers of records & 1057 & 490 & 2988 & 138 & 3407 & 48\\
\addlinespace
Duration in seconds & 390.4 & 176.5 & 996.7 & 49.9 & 1148.6 & 3748.5 \\
\addlinespace
Mean record time (secs) & 0.4 & 0.4 & 0.4 & 0.4 & 0.4 & 78.1 \\
\bottomrule
\end{tabularx}
\end{subtable}

\hspace{\fill}

\begin{subtable}{\textwidth}
\centering
\vspace{20pt}
\subcaption{Number of records and their duration per vocalization class for the validation subset of the \textit{human baby} data set}
\label{tab:effectif_val_bebe}
\begin{tabularx}{\textwidth}{X *{6}c}
\toprule
& \multicolumn{6}{c}{Classes}\\
\cmidrule{2-7}
& Canonical & Crying & Junk & Laughing & Non-Canonical & Noise\\
\midrule
Numbers of records & 386 & 169 & 982 & 51 & 1106 & 16\\
\addlinespace
Duration in seconds & 142.8 & 62.0 & 331.4 & 18.6 & 372.1 & 753.4 \\
\addlinespace
Mean record time (secs) & 0.4 & 0.4 & 0.4 & 0.4 & 0.4 & 47.1 \\
\bottomrule
\end{tabularx}
\end{subtable}

\end{table*}


\begin{table*}

\ContinuedFloat
\caption{Total and per partition distribution of the human baby labeled data set}

\begin{subtable}{\textwidth}
\centering
\vspace{20pt}
\subcaption{Number of records and their duration per class for the testing set of the \textit{human baby} data set}
\label{tab:effectif_test_bebe}
\begin{tabularx}{\textwidth}{X *{6}c}
\toprule
& \multicolumn{6}{c}{Classes}\\
\cmidrule{2-7}
& Canonical & Crying & Junk & Laughing & Non-Canonical & Noise \\
\midrule
Numbers of records & 383 & 164 & 1004 & 52 & 1093 & 17\\
\addlinespace
Duration in seconds & 144.6 & 59.2 & 337.0 & 18.6 & 373.3 & 1772.0 \\
\addlinespace
Mean record time (secs) & 0.4 & 0.4 & 0.4 & 0.4 & 0.4 & 104.2 \\
\bottomrule
\end{tabularx}
\end{subtable}
\end{table*}

\begin{table}[hp]
\caption{Seconds of vocalization for human babies, for each class, over the year}
\begin{tabularx}{\textwidth}{*{5}X}
\toprule \multicolumn{5}{c}{Classes} \\
\cmidrule{1-5}
Canonical & Crying & junk & Laughing & Non-canonical \\
\midrule
3 & 248 & 11 893 & 966 & 113 627 \\
\bottomrule
\end{tabularx}
\label{tab:baby_total_voc_year}
\end{table}

\end{document}